\documentclass[submission,letterpaper, Phys]{SciPost}

\pdfoutput=1
\usepackage{amsmath,amssymb,mathtools,xspace}
\usepackage{booktabs,multirow,graphicx,tabularx,slashed}
\usepackage{hyperref}
\usepackage{color,xcolor}
\usepackage[normalem]{ulem}
\usepackage{enumitem}
\usepackage{feynmp}
\usepackage{braket}
\usepackage{tikz}
\usetikzlibrary{shapes.geometric, arrows}
\usepackage{relsize}

\makeatletter
\@ifundefined{pdfoutput}{}{\DeclareGraphicsRule{*}{mps}{*}{}}
\makeatother

\tikzstyle{generator} = [rectangle, rounded corners, minimum width=3cm, minimum height=1cm,text centered, draw=Gcolor]
\tikzstyle{discriminator} = [rectangle, rounded corners, minimum width=3cm, minimum height=1cm,text centered, draw=Dcolor]
\tikzstyle{io} = [circle, trapezium left angle=70, trapezium right angle=110, minimum width=1cm, minimum height=1cm, text centered, draw=black]

\tikzstyle{process} = [rectangle, minimum width=1cm, minimum height=1cm, text centered, draw=black]
\tikzstyle{decision} = [rectangle, minimum width=1cm, minimum height=1cm, text centered, draw=black]

\tikzstyle{arrow} = [thick,->,>=stealth]
\usepackage{xcolor}

\newcommand{\Langle}{\big\langle}
\newcommand{\Rangle}{\big\rangle}

\newcommand\one{\leavevmode\hbox{\small1\normalsize\kern-.33em1}}

\newcommand{\qqquad}{\qquad \qquad}

\def\slashchar#1{\setbox0=\hbox{$#1$}           
   \dimen0=\wd0                                 
   \setbox1=\hbox{/} \dimen1=\wd1               
   \ifdim\dimen0>\dimen1                        
      \rlap{\hbox to \dimen0{\hfil/\hfil}}      
      #1                                        
   \else                                        
      \rlap{\hbox to \dimen1{\hfil$#1$\hfil}}   
      /                                         
   \fi}

\newcommand{\ie}{\textsl{i.e.}\;}

\newcommand{\madgraph}{\textsc{Madgraph}5\xspace}

\newcommand{\sherpa}{\textsc{Sherpa}\xspace}

\newcommand{\keras}{\textsc{Keras}\xspace}
\newcommand{\tensorflow}{\textsc{TensorFlow}\xspace}

\begin{document}
\begin{fmffile}{feynman}

\begin{center}{\Large \textbf{
How to GAN Event Subtraction
}}\end{center}

\begin{center}
Anja Butter\textsuperscript{1},
Tilman Plehn\textsuperscript{1}, and
Ramon Winterhalder\textsuperscript{1}
\end{center}

\begin{center}
{\bf 1} Institut f\"ur Theoretische Physik, Universit\"at Heidelberg, Germany
winterhalder@thphys.uni-heidelberg.de
\end{center}

\begin{center}
\today
\end{center}


\section*{Abstract}
{\bf Subtracting event samples is a common task in LHC simulation and
  analysis, and standard solutions tend to be inefficient.  We employ
  generative adversarial networks to produce new event samples
  with a phase space distribution corresponding to added or subtracted
  input samples. We first illustrate for a toy example how such a
  network beats the statistical limitations of the training
  data. We then show how such a network can be used to subtract
  background events or to include non-local collinear subtraction
  events at the level of unweighted 4-vector events.}

\vspace{10pt}
\noindent\rule{\textwidth}{1pt}
\tableofcontents\thispagestyle{fancy}
\noindent\rule{\textwidth}{1pt}
\vspace{10pt}

\newpage
\section{Introduction}
\label{sec:intro}

Modern analyses of LHC data are increasingly based on a data-to-data
comparison of measured and simulated events. The theoretical basis of
this approach are generated samples of unweighted or weighted LHC
events.  To match the experimental precision such samples have to be
generated beyond leading order in QCD. In modern approaches to
perturbative QCD at the LHC such simulations include subtraction
terms, leading to events with negative weights. Examples for such
subtraction event samples are subtraction terms for fixed-order real
emission~\cite{Catani:1996jh,Hoche:2018ouj,GehrmannDeRidder:2005cm,Frederix:2008hu,Currie:2013vh},
multi-jet merging including a parton
shower~\cite{Catani:2001cc,Mangano:2002ea}, on-shell
subtraction~\cite{GoncalvesNetto:2012yt}, or the subtraction of
precisely known backgrounds~\cite{Plehn:2011nx}.

Generative adversarial networks or GANs~\cite{goodfellow} are neural
networks which naturally lend themselves to operations on event
samples, as we will show in this paper. Such generative networks have
been proposed for a wide range of tasks related to LHC event
simulation and are expected to lead to significant progress once they
become part of the standard tool box. This includes for instance phase space integration~\cite{bendavid}, event
generation~\cite{dutch,gan_datasets,DijetGAN,Butter:2019cae}, detector
simulations~\cite{calogan1,calogan2,fast_accurate,aachen_wgan1,aachen_wgan2,ATLASShowerGAN,ATLASsimGAN}, 
unfolding~\cite{Datta:2018mwd},
parton
showers~\cite{shower,locationGAN,monkshower,juniprshower,Carrazza:2019cnt}, or searches for physics beyond the Standard Model~\cite{Lin:2019htn}. These references address the general question why
GANs are an useful addition to the standard simulation tool box.
Most recently, we have shown that fully conditional GANs can be used
to invert typical Monte Carlo processes at the LHC, like for instance
a fast detector simulation~\cite{Bellagente:2019uyp}. 

In this paper, we show how GANs can perform simple operations on event
samples, namely adding and subtracting existing samples. Such a
network is trained to generate unweighted events with a phase space
density corresponding to a sum or difference of two or more input
samples. We will illustrate the idea behind a generative event sample
subtraction and addition in Sec.~\ref{sec:toy}. This example shows how
generative networks can beat the statistical limitations of the
training samples. Specifically, we produce events with statistical
fluctuations which are significantly smaller than the corresponding
statistical fluctuations of the training data.  The feature behind
this naively impossible improvement are the excellent interpolation
properties of neural networks in a high-dimensional phase space.

In Sec.~\ref{sec:lhc} we will then subtract unweighted 4-vector events
for the LHC in two examples. First, we subtract the photon continuum
from the complete Drell--Yan process and find the $Z$-pole and the
known interference patterns. This can be seen as a toy example for a
background subtraction at the level of parton-level event samples. For
instance, this setup could allow us to study the kinematics of
four-body decay signals, simulated to high precision from observed
background and signal-plus-background samples.

Finally, we combine a hard matrix element for jet radiation with
collinear subtraction events.  This gives us an event sample that
follows the matrix element minus the subtraction term without any
intermediate binning in the phase space. We show how this subtraction
works even if we do not make use of the local structure of the
subtraction terms. It illustrates how simulations in perturbative QCD
might benefit from GANs, in soft-collinear subtraction, on-shell
subtraction, or a veto-like combination of phase space and parton
shower.

\section{Toy example}
\label{sec:toy}

The advantage of GANs learning how to subtract event samples can be
seen easily from statistical uncertainties in event
counts. Traditionally, we generate the two samples and combine them
through some kind of histogram. If we start with $N + n$ events and
subtract $N \gg n$ statistically independent events, the uncertainty
on the combined events in one bin is given by
\begin{align}
\Delta_n
= \sqrt{ \Delta_{N+n}^2 + \Delta_N^2 } 
\approx \sqrt{2 N} \gg \sqrt{n} \; .
\label{eq:error}
\end{align}
In any bin-wise analysis the bin width has to be optimized. On the one
hand larger bins with more events per bin minimize the relative statistical
error, but on the other hand they reduce the resolution of features.

In our GAN approach we avoid defining such histograms and replace the
explicit event subtraction by a subtraction of interpolated sample
properties over phase space. We will first develop this approach in
terms of a simple toy example and then show how it can be extended to
unweighted 4-vector events as used in LHC simulations. Unfortunately,
there does not (yet) exist a rigid description of statistical and
systematic uncertainties associated with GANs, but we will show how
the fluctuations we observe in our generated samples are visibly
smaller than what we would expect from the input data and
Eq.\eqref{eq:error}.

\subsection{Single subtraction}
\label{sec:toy_sub}

We start with a simple 1-dimensional toy model, \ie toy events which
are described by a single real number $x$. We then define a base distribution $P_B$ and a subtraction
distribution $P_S$ as
\begin{align}
P_B(x)=\frac{1}{x}+0.1
\qquad \text{and} \qquad 
P_S(x)=\frac{1}{x} \;.
\label{eq:diff_sub1a}
\end{align}
The target distribution for the subtraction is then
\begin{align}
P_{B-S} = 0.1 \; .
\label{eq:diff_sub1b}
\end{align}
To produce unweighted subtracted events our GAN is trained to generate
the event sets $\{x_B\}$ and $\{x_S\}$ simultaneously. It thereby
learns the distribution $P_{B-S}$ using the information encoded in the
two input samples.

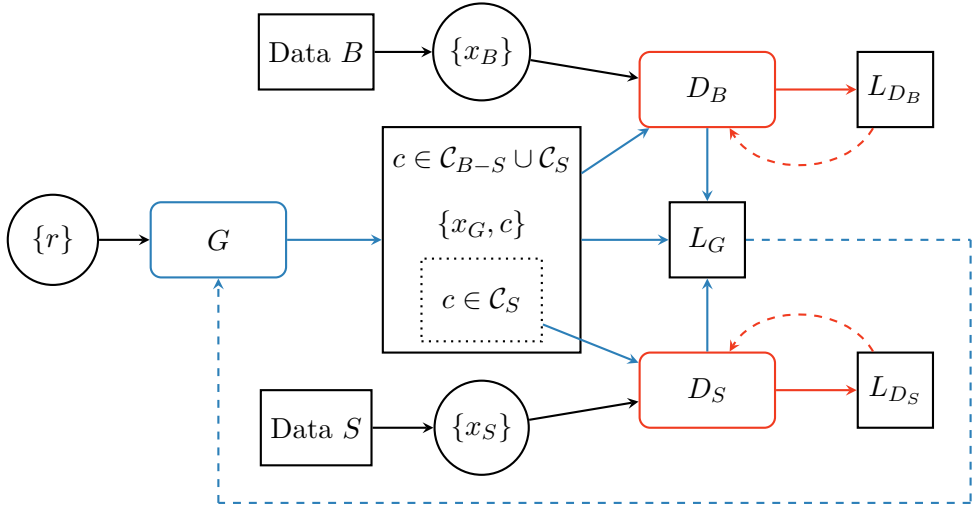
\begin{figure}[t]
\centering
\definecolor{Gcolor}{HTML}{2c7fb8}
\definecolor{Dcolor}{HTML}{f03b20}

\tikzstyle{generator} = [thick, rectangle, rounded corners, minimum width=1.8cm, minimum height=1cm,text centered, draw=Gcolor]
\tikzstyle{discriminator} = [thick, rectangle, rounded corners, minimum width=1.8cm, minimum height=1cm,text centered, draw=Dcolor]
\tikzstyle{io} = [thick,circle, trapezium left angle=70, trapezium right angle=110, minimum width=1.2cm, minimum height=1cm, text centered, draw=black]

\tikzstyle{process} = [thick, rectangle, minimum width=1cm, minimum height=1cm, text centered, draw=black]

\tikzstyle{xG} = [thick,rectangle, minimum width=2.2cm, minimum height=3cm, text depth= 2.2cm, draw=black]
\tikzstyle{s0} = [thick,rectangle, minimum width=2cm, minimum height=3cm, text centered]
\tikzstyle{s1} = [thick, dotted, rectangle, minimum width=1.6cm, minimum height=1.1cm, text centered, draw=black]

\tikzstyle{decision} = [thick,rectangle, minimum width=1cm, minimum height=1cm, text centered, draw=black]

\tikzstyle{dots} = [circle, minimum size=2pt, inner sep=0pt,outer sep=0pt, draw=Dcolor, fill = Dcolor]

\tikzstyle{arrow} = [thick,->,>=stealth]

\begin{tikzpicture}[node distance=2cm]

\node (generator) [generator] {$G$};
\node (random) [io, left of=generator, xshift=-0.2cm, yshift=0cm] {$\{ r \}$};
\draw [arrow, color=black] (random) -- (generator);

\node (xG) [xG, right of=generator, xshift=1.5cm, yshift=0cm] {$c\in\mathcal{C}_{B-S}\cup\mathcal{C}_S$};
\node (xs0) [right of=generator, xshift=1.5cm, yshift=0.2cm] {$\{ x_G,c\}$};
\node (xs1) [s1, right of=generator, xshift=1.5cm, yshift=-0.8cm] {$c\in\mathcal{C}_S$};
\draw [arrow, color=Gcolor] (generator) -- (xG);

\node (d1) [discriminator, right of = xG, xshift=1.0cm, yshift=2cm] {$D_B$};
\node (d2) [discriminator, right of = xG, xshift=1.0cm, yshift=-2cm] {$D_S$};
\node (x1) [io, above of = xG, xshift=0cm, yshift=0.5cm] {$\{x_B\}$};
\node (x2) [io, below of = xG, xshift=0cm, yshift=-0.5cm] {$\{x_S\}$};

\draw [arrow, color=Gcolor] (xG) -- (d1);
\draw [arrow, color=Gcolor] (xs1) -- (d2);
\draw [arrow, color=black] (x1) -- (d1);
\draw [arrow, color=black] (x2) -- (d2);

\node (data1) [process, left of=x1, xshift=-0.2cm, yshift=0cm] {Data $B$};
\node (data2) [process, left of=x2, xshift=-0.2cm, yshift=0cm] {Data $S$};
\draw [arrow, color=black] (data1) -- (x1);
\draw [arrow, color=black] (data2) -- (x2);

\node (dloss1) [process, right of=d1, xshift=0.5cm, yshift=0cm] {$L_{D_B}$};
\node (dloss2) [process, right of=d2, xshift=0.5cm, yshift=0cm] {$L_{D_S}$};
\node (gloss) [process, right of=xG, xshift=1.0cm, yshift=0cm] {$L_{G}$};
\draw [arrow, color=Gcolor] (d1) -- (gloss);
\draw [arrow, color=Gcolor] (d2) -- (gloss);
\draw [arrow, color=Gcolor] (xG) -- (gloss);
\draw [arrow, color=Dcolor] (d1) -- (dloss1);
\draw [arrow, color=Dcolor] (d2) -- (dloss2);
\draw[arrow, dashed, color=Dcolor] (dloss1) [out=-120, in=-60] to (d1);
\draw[arrow, dashed, color=Dcolor] (dloss2) [out=120, in=60] to (d2);

\coordinate[ right of = gloss, xshift=1.5cm, yshift=0.0cm] (out1);
\coordinate[ below of = out1, xshift=0cm, yshift=-1.5cm] (out2);
\coordinate[ below of = generator, xshift=0cm, yshift=-1.5cm] (out3);
\draw[thick, dashed, color=Gcolor] (gloss) -- (out1);
\draw[thick, dashed, color=Gcolor] (out1) -- (out2);
\draw[thick, dashed, color=Gcolor] (out2) -- (out3);
\draw[arrow, dashed, color=Gcolor] (out3) -- (generator);

\end{tikzpicture}
\caption{Structure of our subtraction GAN.  The input $\{r\}$
  describes a batch of random numbers and $\{ x_{B,S} \}$ the true
  input data batches. The label $c$ encodes the category of the
  generated events.  Blue arrows indicate the generator training, red
  arrows the discriminators training.}
\label{fig:GANsimple}
\end{figure}

The corresponding GAN architecture is shown in
Fig.~\ref{fig:GANsimple} and consists of a generator and two
independent discriminators, one for each dataset.  The generator takes
random noise $\{ r\}$ as input and generates samples $\{x_G, c\}$,
where $x_G$ stands for an event and $c$ for a label. The underlying idea is to start from an
event sample which follows $P_B$ and split it into two mutually
exclusive samples following $P_{S}$ and $P_{B-S}$, with class labels
$\mathcal{C}_S$ or $\mathcal{C}_{B-S}$.  During training we demand
that the distribution over events from class $\mathcal{C}_S$ follow
$P_S$ while the full event sample follows $P_B$.  After normalizing
all samples correctly the events with class label $\mathcal{C}_{B-S}$
will then follow the distribution $P_{B-S}$.

Technically, the class label $c$ attached to each event is a real
2-dimensional vector, such that it can be manipulated by the
network. Through a SoftMax function in the final generator layer the
entries of $c$ are forced into the interval $[0, 1]$ and sum to 1. We
then create a so-called one-hot encoding by mapping $c$ to
\begin{align}
c^\text{one-hot}_i = 
\begin{cases}
        1 & \text{if } c_i = \max(c)\\ 
        0 & \text{else } \; .
\end{cases}
\end{align}
This representation is two-dimensional binary and most convenient for
manipulating the samples. We can use it to define the label classes via 
$\mathcal{C}_i =\{c \;|\; c^\text{one-hot}_i =1\}$.\medskip

In Fig.~\ref{fig:GANsimple} we see that for the class $\mathcal{C}_S$ and
the union of $\mathcal{C}_S$ with $\mathcal{C}_{B-S}$ we train
the discriminators to distinguish between events from the input samples
and the generated events.  The training of the discriminators $D_i$
corresponding to the two input samples $\{ x_S \}$ and $\{ x_B \}$ 
uses the standard discriminator loss function for instance in the
conventions of Ref.~\cite{Butter:2019cae}
\begin{align}
L_{D_i}
=   \Langle -\log D_i(x) \Rangle_{x \sim P_T} 
  + \Langle - \log (1-D_i(x)) \Rangle_{x \sim P_G} \; .
\label{eq:dloss1}
\end{align}
We add a regularization and obtain the regularized Jensen-Shannon loss function
\begin{align}
L_{D_i}^\text{(reg)} =
L_{D_i}
+ \lambda_{D_i}
\Langle \left(1- D_i(x)\right)^2 \vert \nabla \phi_i \vert^2 \Rangle_{x \sim P_T} 
+ \lambda_{D_i}
\Langle D_i(x)^2\, \vert \nabla \phi_i \vert^2 \Rangle_{x \sim P_G} \; ,
\label{eq:dloss2}
\end{align}
where we define
\begin{align}
\phi_i(x) = \log \frac{D_i(x)}{1-D_i(x)}\; .
\end{align}
In parallel, we train the generator to fool the discriminators by
minimizing
\begin{align}
L_G
= \sum_i \Langle - \log D_i(x) \Rangle_{x \sim P_G} \; .
\label{eq:gloss1}
\end{align}

An additional aspect in manipulating samples is that we need to keep
track of the normalization or number of events in each class. To
generate a clear and differentiable assignment we introduce the
function
\begin{align}
 f(c) = e^{-\alpha (\max(c)^2-1)^{2\beta}} \in [0,1] 
 \qquad \text{for} \qquad 0\leq c_i \leq 1 \; .
\end{align}
Adapting $\alpha$ and $\beta$ we can make the gradient around the
maximum steeper and push $f(0) \to 0$. In that case $f(c) \approx 1$
only if one of the entries of $c_i \approx 1$. By adding
\begin{align}
L_G^\text{(class)}
&= \left( 1 - \dfrac{1}{b}\sum_{c \in batch} f(c) \right)^2
\label{eq:gloss2}
\end{align}
to the loss function we reward a clear assignment of each event to one
class and generate a clear separation between classes.  Finally, we
use the counting function in combination with masking to fix the
normalization of each sample with
\begin{align}
L_{G_i}^\text{(norm)}
&= \left( 
 \dfrac{\sum_{c \in \mathcal{C}_i} f(c)}
 {\sum_{c \in \mathcal{C}_B} f(c)}
 -\dfrac{\sigma_i}{\sigma_0}
 \right)^2\; .
\label{eq:gloss3}
\end{align}
Adding these losses to the generator loss we get
\begin{align}
	L_G\to L_G^{(\text{full})}= L_G + \lambda_\text{class} L_G^\text{(class)} + \lambda_\text{norm} L_G^\text{(norm)}\; ,
	\label{eq:gloss_full}
\end{align}
with properly chosen factors $\lambda_\text{class}$ and
$\lambda_\text{norm}$. In this paper we always use
$\lambda_\text{class}=\lambda_\text{norm}=1$.  For the denominator in
Eq.\eqref{eq:gloss3} we always choose the approximation of the number
of predicted events in the base class $\mathcal{C}_B = \mathcal{C}_{B-S} \cup \mathcal{C}_S$ as reference value .  The
integrated rates $\sigma_i$ have to be given externally. For our toy
model we can compute them analytically while for an LHC application
they are given by the cross section from the Monte Carlo simulation.

\begin{figure}[t]
\centering
\includegraphics[width=0.49\textwidth]{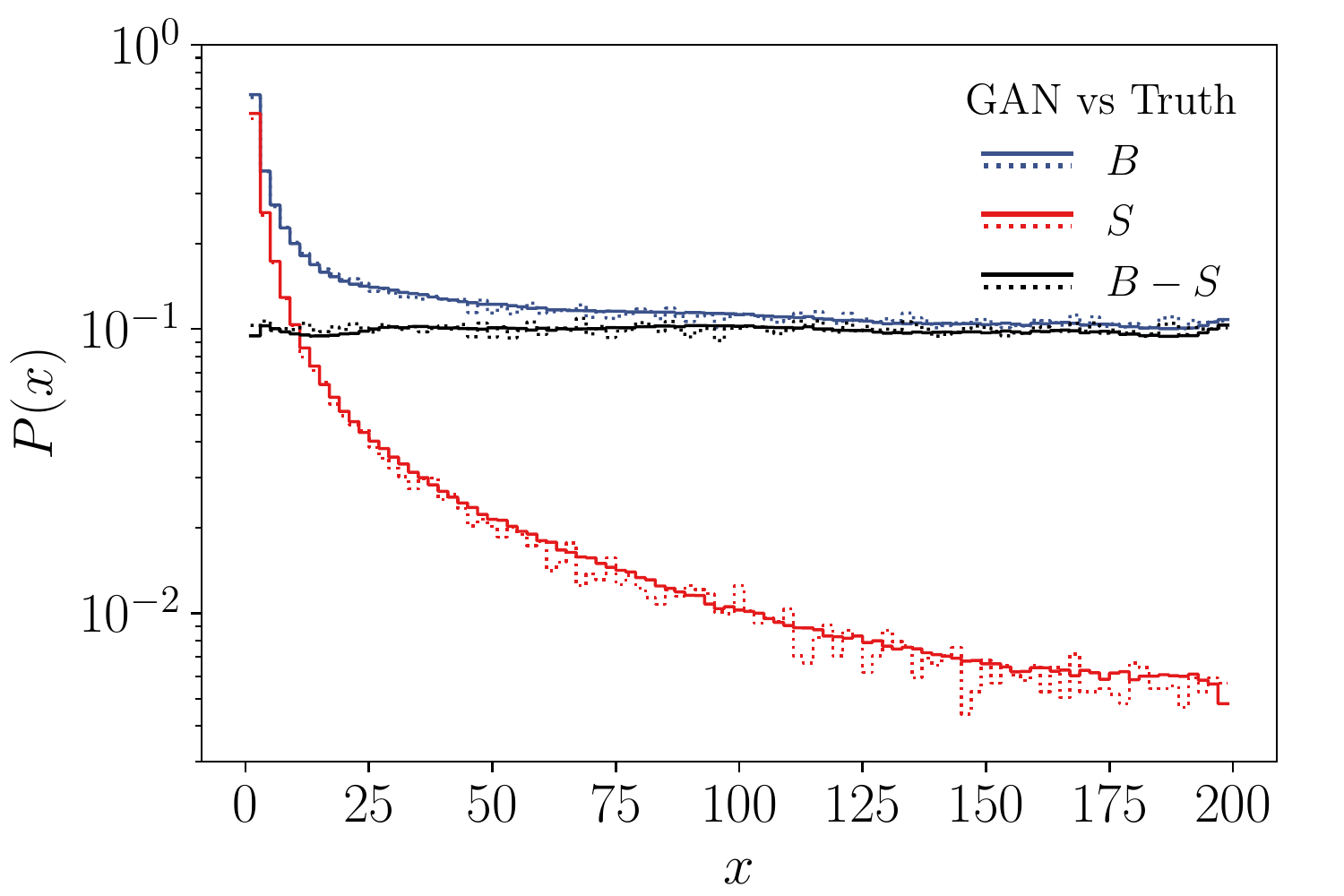}
\includegraphics[width=0.49\textwidth]{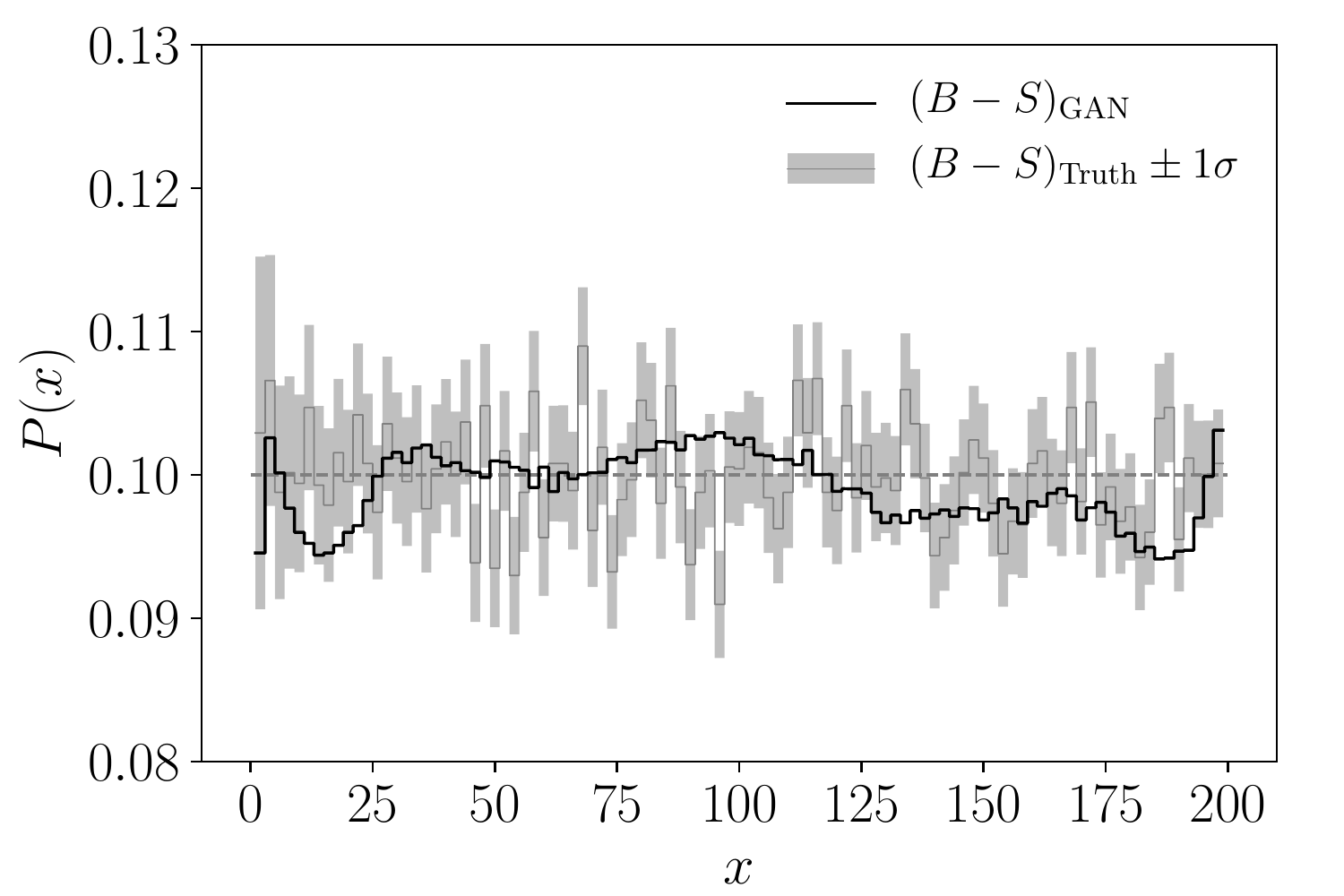}
\caption{Left: Generated (solid) and true (dashed) events for the two
  input distributions and the subtracted output. Right: distribution
  of the subtracted events, true and generated, including the error
  envelope propagated from the input statistics.}
\label{fig:toy_sub}
\end{figure}

Our GAN uses a vector of random numbers as input. The size of the
vector has to be at least the number of degrees of freedom.  For the
implementation we use \keras~2.2.4~\cite{keras} with a
\tensorflow~1.14 back-end~\cite{tensorflow}. The discriminator and
generator networks consist of 5 layers with 128 units per layer using
the ELU activation function. With $\lambda_{D_i}=5 \cdot 10^{-5}$ and
a batch size of 1024 events, we run for 4000 epochs. Each epoch
consists of one update of the generator and 20 updates of the
discriminator. We found that the intense training of the discriminator
is necessary to reach sufficiently precise results. To obtain a good
separation of the classes with $f(c)$ we set $\alpha = 10$ and $\beta
= 1$.  Finally, using the \textsc{Adam}~\cite{adam} optimizer throughout this
paper, we choose a learning rate of $3 \cdot 10^{-4}$ for generator
and discriminator and a large decay of the learning rate of $2 \cdot
10^{-2}$ for the discriminator which stabilizes the training.  The
decay for the generator is slightly smaller with $5 \cdot
10^{-3}$. Our training datasets consist of $10^5$ samples for each
dataset $\{ x_S \}$ and $\{ x_B \}$.\medskip

We show numerical results for a single GAN subtraction and analyze the
size of the statistical fluctuations in Fig.~\ref{fig:toy_sub}. In the
left panel we show the two input distributions defined in
Eq.\eqref{eq:diff_sub1a}, as well as the true and generated subtracted
distribution. The dotted lines illustrate the shape of the training
dataset, while the full lines show the generated distribution using
$5\cdot10^6$ events. The former two distributions only serve to
confirm that the GAN learns the input information correctly. The
generated subtracted events indeed follow the probability distribution
in Eq.\eqref{eq:diff_sub1b}. Aside from the fact that all three
distributions show excellent agreement between truth and GANned
events, we see how the neural network interpolates especially in the
tail of the distribution.  In the right panel of
Fig.~\ref{fig:toy_sub} we zoom into the subtracted sample to compare
the statistical uncertainties from the input data with the behavior of
the GAN.  The uncertainty is estimated from the number of events per
bin in the base and subtraction histogram $N_B$ and $N_S$, taking into
account the corresponding normalization factors $n_B$ and $n_S$. In
analogy to Eq.\eqref{eq:error} we compute it as
\begin{align}
\begin{split}
\Delta_{B-S}
&=\Delta_{n_B N_{B}-n_S N_{S}}\\
&= \sqrt{ \Delta_{n_B N_B}^2 + \Delta_{n_S N_S}^2 } \\
&= \sqrt{n_B^2 N_B + n_S^2 N_S } \; .
\end{split}
\end{align}
As mentioned above, we expect the GAN to deliver more stable results
than we could expect from the input sample, because the GAN
interpolates all input distributions.  This way we avoid a bin-by-bin
statistical uncertainty of the subtracted sample.  Indeed, our
subtracted curve in the right panel of Fig.~\ref{fig:toy_sub} lies
safely within the $1\sigma$ region of the data. The statistical
fluctuations of the GANned events are much smaller than the
statistical fluctuations in the input data. On the other hand, the
GANned distribution shows systematic deviations, but also at a visibly
smaller level than the statistical fluctuation of the input
data. While this observation does not imply a proof that GANs can beat
the statistical limitations of the input data, they give a clear hint
that the interpolation properties can balance statistics at some
level.

\subsection{Combined subtraction and addition}
\label{sec:toy_add}

\begin{table}[b!]
\begin{small} \begin{center}
\begin{tabular}{l c c c c c c c c}
\toprule
& $\mathcal{C}_{B-S}$ & $\mathcal{C}_S$ & $\mathcal{C}_A$\\
\midrule
Data $B$ & $1$ & $1$ & $0$\\
Data $S$ &  $0$ & $1$ & $0$\\
Data $A$ &  $0$ & $0$ & $1$\\
\midrule
$B-S+A$ & $1$ & $0$ & $1$\\
\bottomrule
\end{tabular}
\end{center} \end{small}
\caption{Category assignment for a combined addition and subtraction
  of three samples.}
\label{tab:classes_small}
\end{table}

To show how our approach could be generalized to subtracting and
adding any number of event samples we can extend our single
subtraction toy model by a third sample to be added to the difference
described in Eq.\eqref{eq:diff_sub1b}. We now consider three samples
corresponding to the 1-dimensional distributions
\begin{align}
P_B(x)=\frac{1}{x}+0.1
\qqquad
P_S(x)=\frac{1}{x}
\qqquad 
P_A(x)=\frac{m}{\pi}\frac{\gamma}{\gamma^2+(x-x_0)^2} \;.
\label{eq:diff_sub2a}
\end{align}
As a third input we add the Breit-Wigner distribution $P_A$, so our
target distribution becomes
\begin{align}
P_{B-S+A}
&=\frac{m}{\pi}\frac{\gamma}{\gamma^2+(x-x_0)^2}+0.1  \notag \\
&=\frac{5}{\pi}\frac{10}{100+(x-90)^2}+0.1 \; ,
\label{eq:diff_sub2b}
\end{align}
for the values $m=5$, $\gamma = 10$, and $x_0 = 90$.  We now sample
$\{x_B\}$, $\{x_S\}$ and $\{x_A\}$ individually from the input
distributions and want to learn the probability distribution
$P_{B-S+A}$. The approach is the same as described before, but for
three classes as shown in Tab.~\ref{tab:classes_small} and a
three-dimensional class vector. Treating the subtraction exactly as
before we obtain our target distribution $P_{B-S+A}$ by adding the
event with class $\mathcal{C}_A$.

Compared to the sample subtraction introduced before, adding samples
is obviously not a big challenge. In principle, we could just add the
unweighted event samples in the correct proportion, learn the phase
space structure with a GAN, and then generate any number of events
very efficiently. The reason why we discuss this aspect here is that
it shows how our subtraction GAN can be generalized easily.

\begin{figure}[t]
\centering
\includegraphics[width=0.49\textwidth]{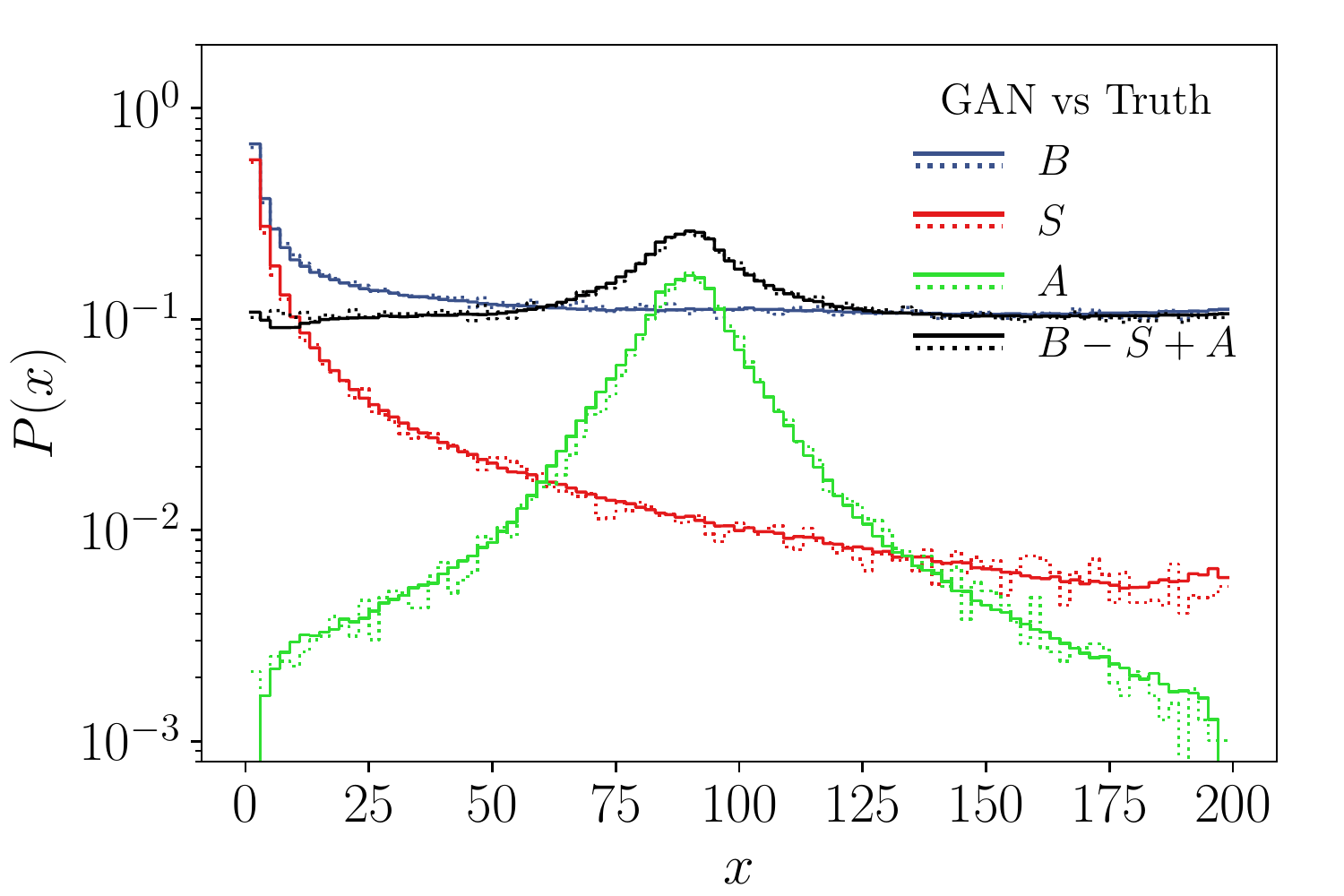}
\includegraphics[width=0.49\textwidth]{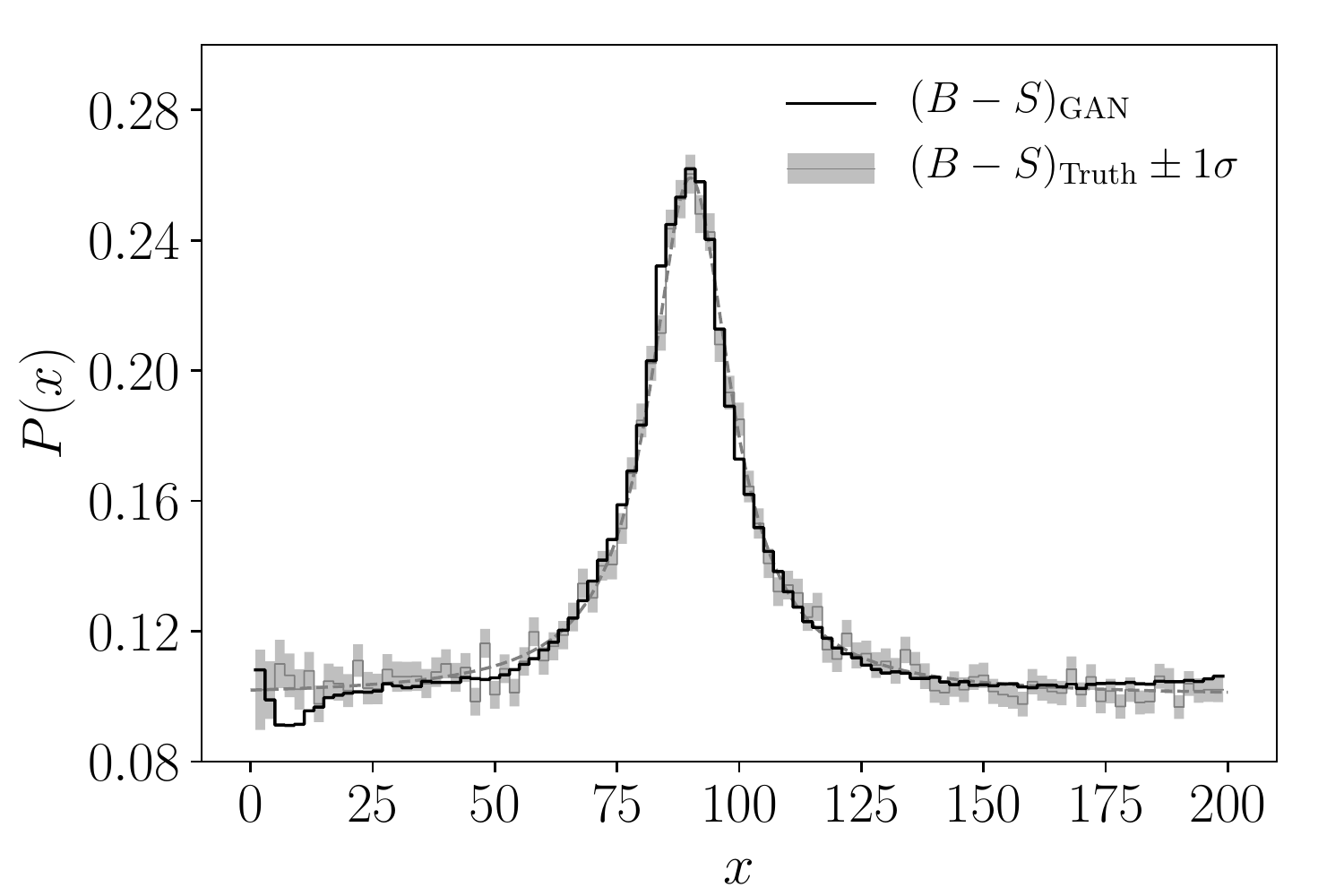}
\caption{Left: Generated (solid) and true (dotted) events for the
  three input distributions and the combined output. Right:
  distribution of the combined events, true and generated, including
  the error envelope propagated from the input statistics.}
\label{fig:toy_add}
\end{figure}

In Fig.~\ref{fig:toy_add} we show the numerical results of subtracting
one distribution $\{ x_S \}$ from the base distribution $\{ x_B \}$
and adding a second distribution $\{ x_A \}$ with a distinct
feature. As before, this combination is learned from the three input
distributions without binning the corresponding phase space. The
hyper-parameters are slightly modified with respect to the simple
subtraction model.  The networks now consist of 7 layers with 128
units which we train for 1000 epochs with 4 iterations.  We fix the
relative weight of the gradient penalty to $\lambda_{D_i}=5 \cdot
10^{-5}$.  The separation of the three classes is efficient for
$\alpha = 5$ and $\beta = 1$.  Finally, we set the learning rate to $8
\cdot 10^{-4}$ and its decay to $2 \cdot 10^{-2}$ for generator and
discriminator.  The remaining parameters are the same as for the pure
subtraction case.  In the left panel of Fig.~\ref{fig:toy_add} we
confirm that the GAN indeed learns the three input structures
correctly and interpolates each of them smoothly. We also see that the
generated events follow the combination $B-S+A$ with its flat tails
and the central Breit--Wigner shape. As for the pure subtraction in
Fig.~\ref{fig:toy_sub} we also compare the statistical fluctuation of
the binned input data with the behavior of the GANned events. The GAN
extracts the additional Breit--Wigner feature with high precision,
but, as always, some systematic deviations arise in the tails of the
distribution.

\subsection{General setup}
\label{sec:toy_gen}

\begin{figure}[t]
\centering

\definecolor{Gcolor}{HTML}{2c7fb8}
\definecolor{Dcolor}{HTML}{f03b20}

\tikzset{fontscale/.style = {font=\relsize{#1}}
    }

\tikzstyle{generator} = [thick, circle, rounded corners, minimum width=2.0cm, minimum height=1cm,text centered, draw=Gcolor, fontscale=1.5]
\tikzstyle{discriminator} = [thick, circle, rounded corners, minimum width=2.0cm, minimum height=1cm,text centered, draw=Dcolor, fontscale=1.5]
\tikzstyle{io} = [thick,circle, trapezium left angle=70, trapezium right angle=110, minimum width=1.8cm, minimum height=1cm, text centered, draw=black, fontscale=1.5]

\tikzstyle{process} = [rectangle, minimum width=1cm, minimum height=1cm, text centered, draw=black]
\tikzstyle{decision} = [rectangle, minimum width=1cm, minimum height=1cm, text centered, draw=black]

\tikzstyle{dots} = [circle, minimum size=2pt, inner sep=0pt,outer sep=0pt, draw=Dcolor, fill = Dcolor]

\tikzstyle{arrow} = [thick,->,>=stealth]

\begin{tikzpicture}[node distance=2cm, scale=0.57 , every node/.style={scale=0.57}]

\node (inG) [io] {$\{ x_G, c\}$};
\node (generator) [generator, left of = inG, xshift=-1.5cm] {$G$};
\node (random) [io,left of=generator, xshift=-.2cm, yshift=-1.6cm] {$\{ r \}$};
\node (gloss) [io,left of=generator, xshift=-2.0cm, yshift=0cm] {$L_G$};
\draw [arrow, color=black] (random) -- (generator);
\draw [arrow, color=black] (generator) -- (inG);
\draw[arrow,dashed, color=Gcolor] (gloss) -- (generator);
\draw[arrow,color=Gcolor, bend right] (inG) to [auto] (gloss);

\node (d1) [discriminator, left of = inG, xshift=0.25cm, yshift=3.03cm] {$D_B$};
\node (x1) [io,left of=inG, xshift=-2.098cm, yshift=4.719cm] {$\{ x_B \}$};
\node (dloss1) [io,left of=inG, xshift=-0.038cm, yshift=5.908cm] {$L_{D_B}$};

\draw [arrow, color=Dcolor] (x1) -- (d1);
\draw [arrow, color=black] (inG) --node[scale=0.8, sloped, anchor=center, above, color=black, rotate=0, fontscale=1.2]{$c\in \bigcup\limits^M\mathcal{C}_i$}  (d1);
\draw[arrow, color=Dcolor] (d1) -- (dloss1);
\draw[arrow,dashed, color=Dcolor, bend left] (dloss1) [auto] to (d1);

\coordinate[left of=inG, xshift=-1.75cm, yshift=6.495cm] (gloss1) ;
\draw[arrow, color=Gcolor] (d1) -- (gloss1);

\node (d2) [discriminator, left of = inG, xshift=3.75cm, yshift=3.03cm] {$D_{S_1}$};
\node (x2) [io,left of=inG, xshift=4.038cm, yshift=5.908cm] {$\{ x_{S_1} \}$};
\node (dloss2) [io,left of=inG, xshift=6.098cm, yshift=4.719cm] {$L_{D_{S_1}}$};

\draw [arrow, color=Dcolor] (x2) -- (d2);
\draw [arrow, color=black] (inG) -- node[scale=0.8, sloped, anchor=center, above, color=black, fontscale=1.2]{$c\in \mathcal{C}_1$}  (d2);
\draw[arrow, color=Dcolor] (d2) -- (dloss2);
\draw[arrow,dashed, color=Dcolor, bend left] (dloss2) [auto] to (d2);

\coordinate[left of=inG, xshift=5.75cm, yshift=6.495cm] (gloss2) ;
\draw[arrow, color=Gcolor] (d2) -- (gloss2);

\node (d3) [discriminator, right of = inG, xshift=1.5cm] {$D_{S_2}$};
\node (x3) [io,left of=inG, xshift=8.136cm, yshift=1.189cm] {$\{ x_{S_2} \}$};
\node (dloss3) [io,left of=inG, xshift=8.136cm, yshift=-1.189cm] {$L_{D_{S_2}}$};

\draw [arrow, color=Dcolor] (x3) -- (d3);
\draw [arrow, color=black] (inG) -- node[scale=0.8, sloped, anchor=center, above, color=black, fontscale=1.2]{$c\in \mathcal{C}_2$} (d3);
\draw[arrow, color=Dcolor] (d3) -- (dloss3);
\draw[arrow,dashed, color=Dcolor, bend left] (dloss3) [auto] to (d3);

\coordinate[left of=inG, xshift=9.5cm] (gloss3) ;
\draw[arrow, color=Gcolor] (d3) -- (gloss3);

\node (dN) [discriminator, left of = inG, xshift=0.25cm, yshift=-3.03cm] {$D_{A_N}$};
\node (xN) [io,left of=inG, xshift=-.038cm, yshift=-5.908cm] {$\{ x_{A_N} \}$};
\node (dlossN) [io,left of=inG, xshift=-2.098cm, yshift=-4.719cm] {$L_{D_{A_N}}$};

\draw [arrow, color=Dcolor] (xN) -- (dN);
\draw [arrow, color=black] (inG) -- node[scale=0.8, sloped, anchor=center, above, color=black, fontscale=1.2]{$c\in \mathcal{C}_{M+N}$} (dN);
\draw[arrow,dashed, color=Dcolor] (dlossN) -- (dN);
\draw[arrow, color=Dcolor, bend right] (dN) [auto] to (dlossN);

\coordinate[left of=inG, xshift=-1.75cm, yshift=-6.495cm] (glossN) ;
\draw[arrow, color=Gcolor] (dN) -- (glossN);

\coordinate[dots, left of = inG, xshift=3.75cm, yshift=-3.031cm] (dot1);
\coordinate[dots, left of = inG, xshift=2.928cm, yshift=-3.375cm] (dot2);
\coordinate[dots, left of = inG, xshift=4.459cm, yshift=-2.491cm] (dot3);

\coordinate[dots, left of = inG, xshift=2.0459cm, yshift=-3.4997cm] (dot6);
\coordinate[dots, left of = inG, xshift=5.0079cm, yshift=-1.7896cm] (dot7);

\coordinate[ left of = inG, xshift=2.098cm, yshift=-7.499cm] (out1);
\coordinate[ left of = inG, xshift=8.445cm, yshift=-3.835cm] (out2);

\draw[arrow, color=Gcolor] (out2) arc[radius=7.5cm, start angle=-30.7519, end angle=173]  (gloss);
\draw[arrow, color=Gcolor] (out1) arc[radius=7.5cm, start angle=-89.2481, end angle=-173]  (gloss);
\draw[thick, dash pattern=on 1pt off 14pt, color=Gcolor] (out1) arc[radius=7.5cm, start angle=-89.2481, end angle=-30.7519] to (out2);

\end{tikzpicture}
\caption{Structure of our general subtraction and addition GAN. The
  input $\{r\}$ describes a batch of random numbers and $\{ x \}$ the
  true input data or generated batches. The label $c$ encodes the
  category of the generated events.  Blue arrows indicate the
  generator training, red arrows the discriminators training.}
\label{fig:GANgeneral}
\end{figure}
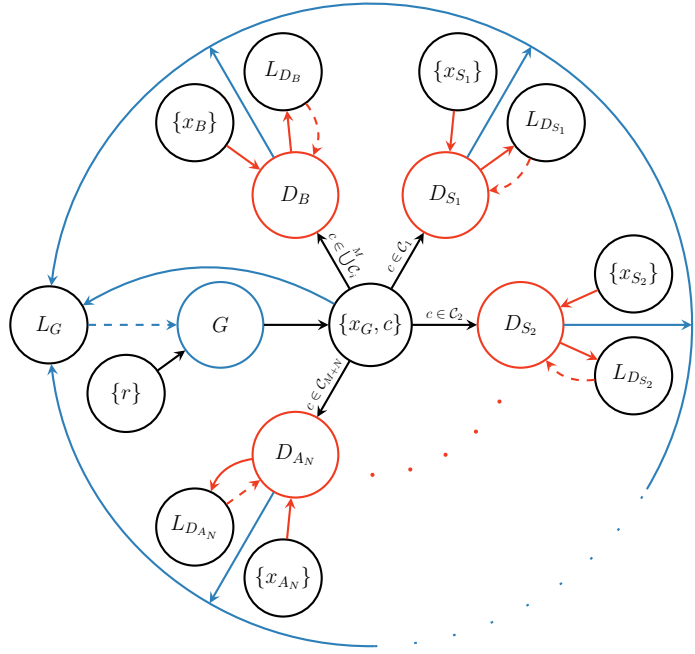

\begin{table}[b!]
\begin{small} \begin{center}
\begin{tabular}{l c c c c c c c c}
\toprule
& $\mathcal{C}_0$ & $\mathcal{C}_1$ & $\mathcal{C}_2$ & $\cdots$ & $\mathcal{C}_M$ & $\mathcal{C}_{M+1}$ & $\cdots$ & $\mathcal{C}_{M+N}$\\
\midrule
Data $B$ & $1$ & $1$ & $1$  & $\cdots$ & $1$ & $0$  & $\cdots$ & $0$\\
Data $S_1$ &  $0$ & $1$ & $0$  & $\cdots$ & $0$ & $0$  & $\cdots$ & $0$\\
Data $S_2$ &  $0$ & $0$ & $1$  &  & $0$ & $0$  & $\cdots$ & $0$\\
$\vdots$ & $\vdots$ & $\vdots$ &  & $\ddots$ & & $\vdots$ & & $\vdots$\\
Data $S_M$ &  $0$ & $0$ & $0$  &  & $1$ & $0$  & $\cdots$ & $0$\\
Data $A_1$&  $0$ & $0$ & $0$  & $\cdots$ & $0$ & $1$  &  & $0$\\
$\vdots$ & $\vdots$ & $\vdots$ & $\vdots$ & $\ddots$ & $\vdots$ & & $\ddots$ \\
Data $A_N$ & $0$ & $0$ & $0$  & $\cdots$ & $0$ & $0$  &  & $1$\\
\midrule
Combination & $1$ & $0$ & $0$  & $\cdots$ & $0$ & $1$ & $\cdots$ & $1$\\
\bottomrule
\end{tabular}
\end{center} \end{small}
\caption{Details for the category selection in the general case.}
\label{tab:details}
\end{table}

Finally, we note that our network setup is not limited to three
classes. We can generalize it to a base distribution, $M$ subtraction
datasets, and $N$ added datasets. The corresponding category
assignment, generalized from Tab.\ref{tab:classes_small}, is given in
Tab.~\ref{tab:details} and encoded in an enlarged classification
vector $c$. The base class is then defined as
\begin{align}
\mathcal{C}=\bigcup\limits_{i=0}^M C_i\; .
\end{align}
In this case the network has to learn all $M+N+1$ input
distributions through individual discriminators $D_B$, $D_{S_i}$, and
$D_{A_j}$ with $i\le M$ and $j \le N$. The rough structure of the
network is given in Fig.~\ref{fig:GANgeneral}. The training of the
generator follows directly from the description above. While we do not
benchmark this extended setup in this paper, we expect it to be useful
when a set of subtraction terms accounts for different features, and
splitting them improves their simulation properties.

Until now we have always assumed that we can subtract a sample $\{ x_S
\}$ from a sample $\{ x_B \}$ and find a well-behaved distribution for
$B-S$. Specifically, the resulting probability $P_{B-S}$ should be
positive all over phase space. This is not always the case. First, we
note that a global sign of the combination is not a problem, because
we can always learn $S-B$ instead of $B-S$. Next, changing signs in
the $S$ or $B$ contributions can be accommodated by splitting the
respective sample according to the sign and applying the combined
subtraction and addition described in Sec.~\ref{sec:toy_add}.  A
phase-space dependent sign in $S-B$ could be most easily accommodated
by adding a constant off-set either by hand or again using the
combined subtraction and addition. A typical example would be to add
the Born term to the virtual correction before subtracting the dipole.

In cases where this is not a suitable solution, we can replace the
categories $\mathcal{C}_{B-S}$ and $\mathcal{C}_S$ by the three
categories $\mathcal{C}_{B \cap S}$, $\mathcal{C}_{B \setminus S}$,
and $\mathcal{C}_{S \setminus B}$. They indicate events corresponding
to $B$ and $S$, only $B$, or only $S$. The discriminator compares for
instance the combination of $\mathcal{C}_{B \cap S}$ and
$\mathcal{C}_{B \setminus S}$ with the $B$-data. The difference $B-S$
will be given by events with label $\mathcal{C}_{B \setminus S}$ in
regions where $B>S$ and events with label $\mathcal{C}_{S \setminus B}$ in
regions where $B<S$, the latter weighted with weight minus one.  While
this simple extension of the label vector is very straightforward, the
third category induces an additional degree of freedom in the way the
network can distribute events into different categories. This freedom
needs to be constrained to prevent the network from simply assigning
for instance all events into the categories $\mathcal{C}_{B \setminus
  S}$, and $\mathcal{C}_{S \setminus B}$. A possible solution would be
to maximize the number of events in $\mathcal{C}_{B \cap S}$ via a
term in the loss function and force the network to share as many
events between the distributions as possible.

\section{LHC events}
\label{sec:lhc}

After showing how it is possible to GAN-subtract 1-dimensional event
samples from each other we have to show how such a tool can be applied
in LHC physics. In this case the (unweighted) events are 4-momenta of external particles. We ignore all information on the
particle identification, except for its mass, which allows us to
reduce external 4-momenta to external 3-momenta~\cite{Butter:2019cae,Bellagente:2019uyp}. Because the input
events might have been object to detector effects we do not assume
energy-momentum conservation for the entire event. This means that the
network has to learn the 4-dimensional energy-momentum conservation
and this subtraction of simple LHC events is inherently
multi-dimensional. We will present two simple examples for LHC event
subtraction, the separation of on-shell photon and $Z$ contributions
to the Drell-Yan process and the subtraction of collinear gluon
radiation in $Z$+jet production.

\subsection{Background subtraction}
\label{sec:lhc_back}

\begin{figure}[b!]
\begin{center}
\input{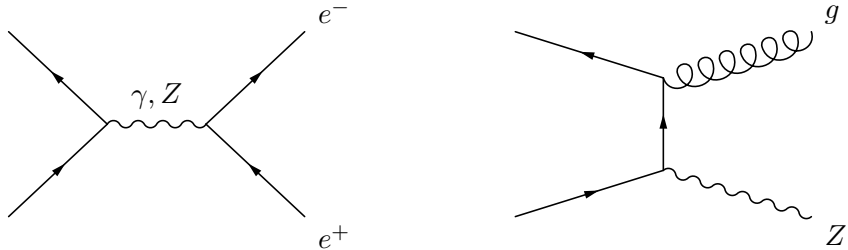}
\end{center}
\caption{Sample Feynman diagrams for the background subtraction (left) and
  collinear subtraction (right) applications.}
\label{fig:feyn}
\end{figure}

\begin{figure}[t]
\centering
\includegraphics[page=1, width=0.49\textwidth]{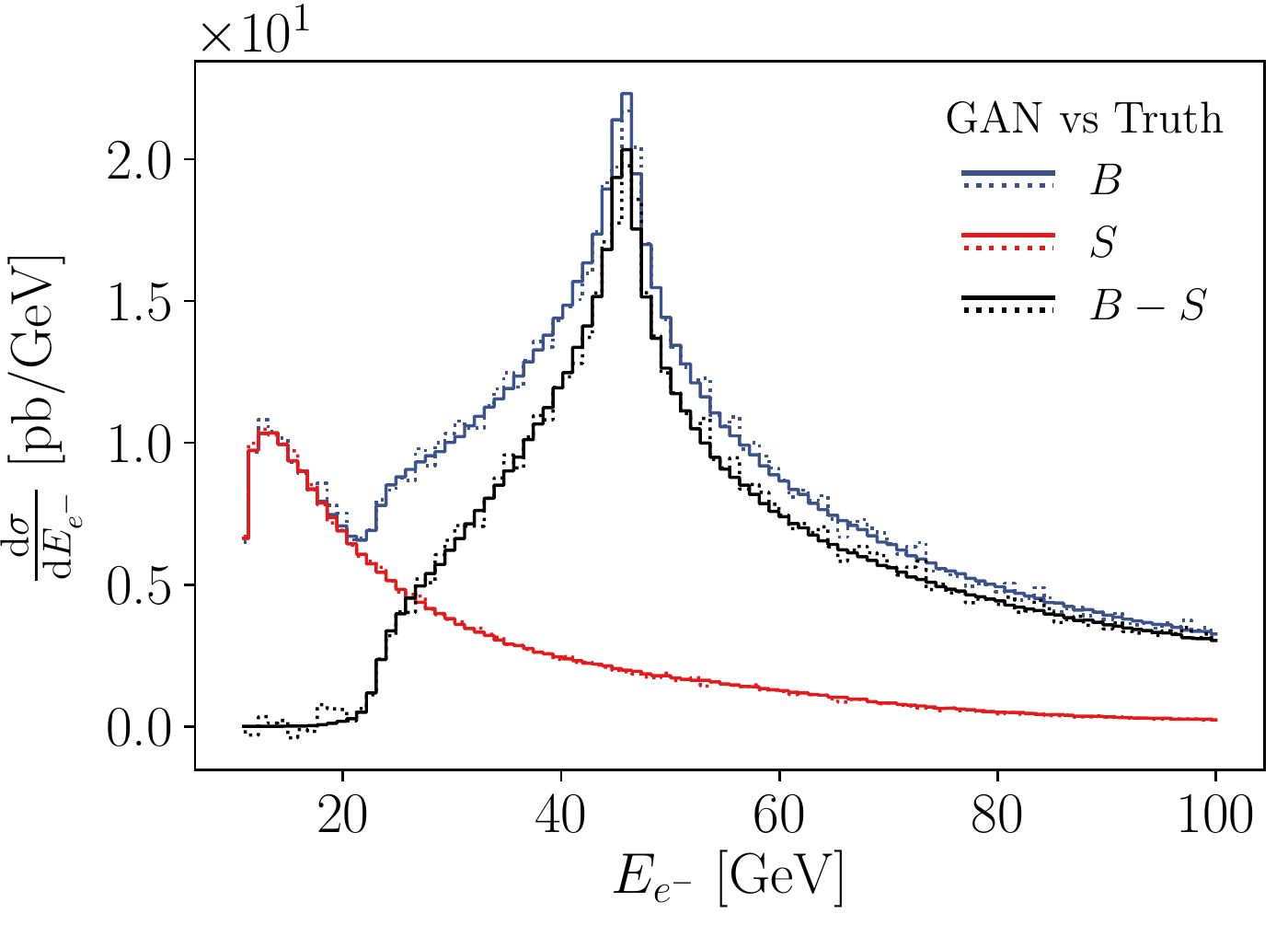}
\includegraphics[page=1, width=0.49\textwidth]{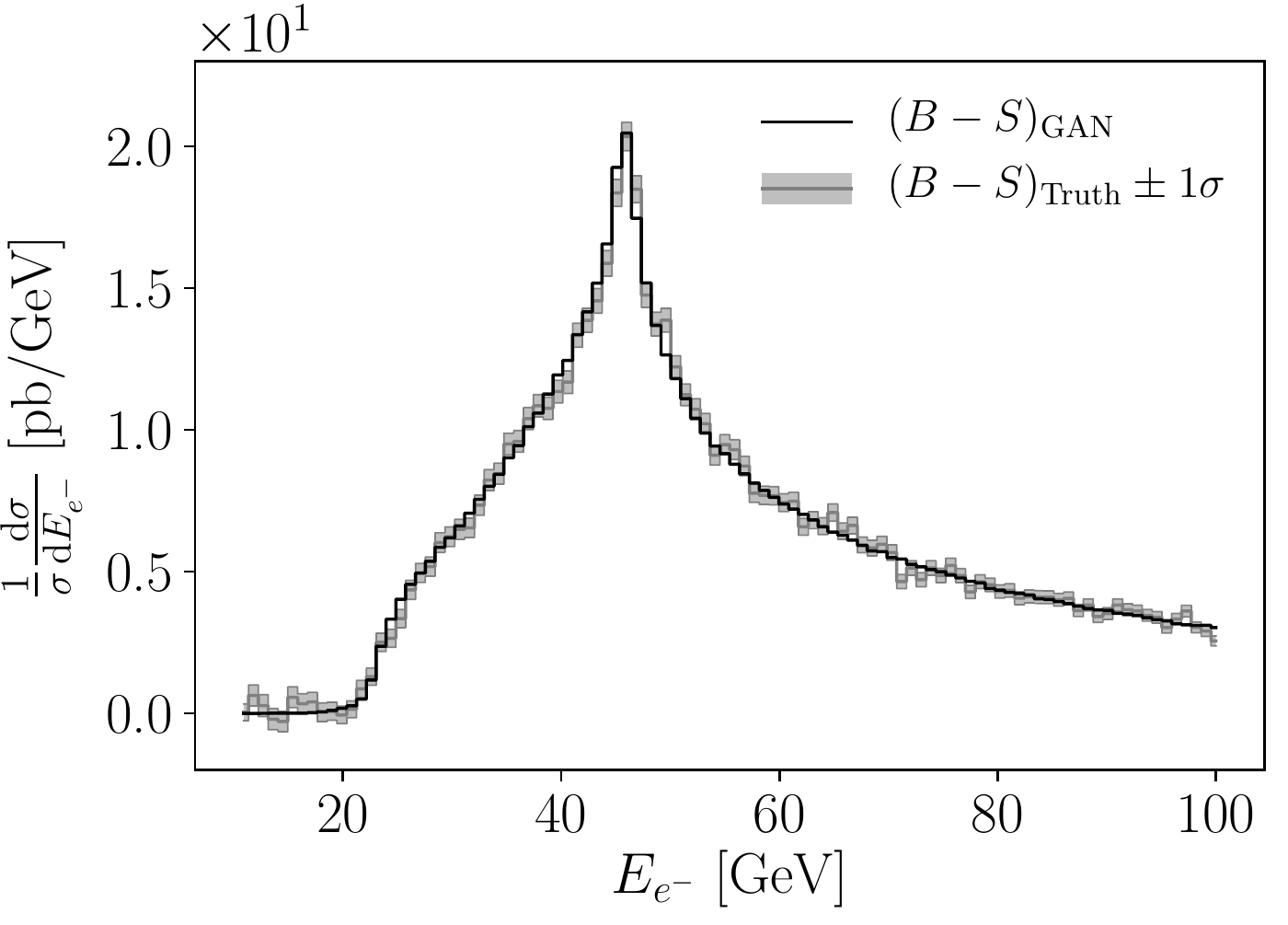}\\
\includegraphics[page=2, width=0.49\textwidth]{figures/lhc_dist}
\includegraphics[page=2, width=0.49\textwidth]{figures/lhc_dist_zoom}
\caption{Left: Generated (solid) and true (dashed) $e^+ e^-$ events at
  the LHC for the two input distributions and the subtracted
  output. Right: distribution of the subtracted events, true and
  generated, including the error envelope propagated from the input
  statistics.}
\label{fig:lhc_sub1}
\end{figure}

Our first example for event subtraction at the LHC is the Drell--Yan
process, which receives contributions with distinct phase space
features from the photon and from the $Z$-boson, as seen in
Fig.~\ref{fig:feyn}. The specific question in our setup is if we can
subtract a background-like photon continuum contribution from the full
process and generate events only for the $Z$-exchange combined with
the interference term,
\begin{align}
\begin{split}
B: \qquad pp &\to  e^+ e^- \\
S: \qquad pp &\to \gamma \to e^+ e^- \; .
\end{split}
\end{align}
We generate 1M events with \madgraph~\cite{madgraph} for an LHC energy of
13~TeV, applying minimal cuts on the outgoing electrons. We require a 
minimal $p_T$ of 10~GeV, a maximal rapidity of 2.5 for each electron,
and a minimal angular separation of 0.4.
We do not apply a detector simulation at this stage, because our focus is
on comparing the generated and true distributions, and we have shown
that detector simulations can be included trivially in our GAN
setup~~\cite{Butter:2019cae,Bellagente:2019uyp}.

Aside from the increased dimension of the phase space the subtraction
GAN has exactly the same structure as the toy example of
Sec.~\ref{sec:toy}. The hyper-parameters have to be adjusted to the
increased dimensionality of the phase space.  We use a 16-dimensional
latent space. The discriminator and generator networks consist of 8
layers with 80 and 160 units per layer, respectively. In this
high-dimensional case we use the LeakyRelu activation function.
Further, we choose $\lambda_{D_i}=10^{-5}$ and a batch size of 1024
events and train for 1000 epochs. Each epoch consists of 5 iterations
in which the discriminator gets updated twice as much as the
generator. For a proper separation of the classes with $f(c)$ we set
$\alpha = 5$ and $\beta = 1$. Finally, we choose a large decay of the
learning rate of $10^{-2}$ which stabilizes the training and pick a
learning rate at the beginning of $10^{-3}$ .  Our training datasets
consist of $10^5$ samples for each dataset $\{ x_B \}$ and $\{ x_S
\}$.

In Fig.~\ref{fig:lhc_sub1} we show the performance of the LHC event
subtraction for two example distributions. First, we clearly see the
$Z$-mass peak in the lepton energy of the full sample, compared with
the feature-less photon continuum in the subtraction sample.  The
subtracted curve is expected to describe the $Z$-contribution and the
interference. It smoothly approaches zero for small lepton energies,
where the interference is negligible. Above that we see the Jacobian
peak from the on-shell decay, and for larger energies a small
interference term enhancing the high-energy tail. In the (usual) right
panel we show the subtracted curve including the statistical
uncertainties from the input samples. As the second observable we show
the transverse momentum of the electron. Here the $Z$-pole appears as
a softened endpoint at $m_Z/2$. The photon continuum dominates the
combined distribution for small transverse momenta. Indeed, the
GAN-subtracted on-shell and interference contribution is localized
around the endpoint, with a minor shift in the resolution at the edge.

Obviously, our subtraction of the background to a di-electron
resonance is not a state-of-the-art problem in LHC physics. A more
interesting application of our method could be four-body decays. We
could start from a combined signal plus background sample of Higgs
decays to four fermions, generate a background-only sample using
control regions, and then GAN a set of signal events. While in a
regular analysis the events we obtain from subtracting a background
from the signal-plus-background sample do not reflect the signal
properties, our GANned subtraction events should reflect all kinematic
features of the signal events in the data. As an example we would like to mention the sPlot approach~\cite{sPlot, sPlot2} which can be  used to statistically subtract background contributions.

\subsection{Collinear subtraction}
\label{sec:lhc_coll}

The second example for event subtraction at the LHC is collinear
radiation off the initial state, for instance
\begin{align}
\begin{split}
B: \qquad pp &\to  Z g \quad \text{(matrix element)} \\
S: \qquad pp &\to  Z g \quad \text{(collinear approximation)} \; 
\end{split}
\end{align}
We generate 1M events for the hard process with
\sherpa~\cite{Bothmann:2019yzt}, where the $Z$-boson decays to
electrons. For the network we combine the electron and positron
momenta to a 4-momentum of the $Z$-boson, so we obtain a Breit--Wigner
distribution with $m_{ee} = 66~...~116$~GeV instead of an on-shell
condition. We then subtract the corresponding Catani-Seymour
dipoles~\cite{Catani:1996jh} for the gluon radiation off each of the
incoming quarks, based on 1M events each. The corresponding Feynman
diagram is shown in the right panel of Fig.~\ref{fig:feyn}. To avoid
the soft divergence we require $p_{T,g} > 1$~GeV in the training data, a
smaller cutoff would be possible but increase the training time. We
apply the same external cutoff to the GANned samples, aligning the
phase space boundaries of the training and GANned data sets by hand.

\begin{figure}[t]
\centering
\includegraphics[page=1, width=0.49\textwidth]{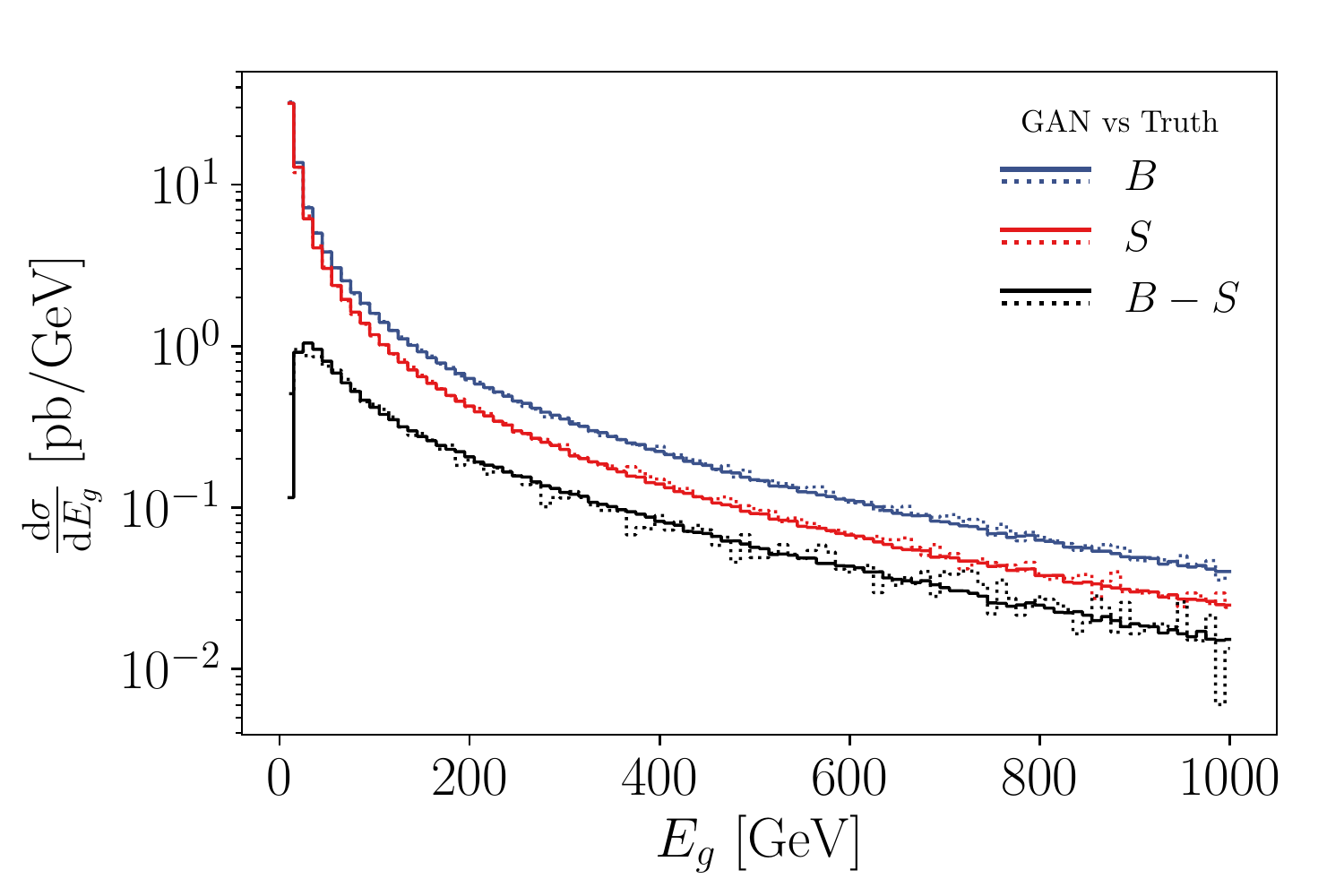}
\includegraphics[page=1, width=0.49\textwidth]{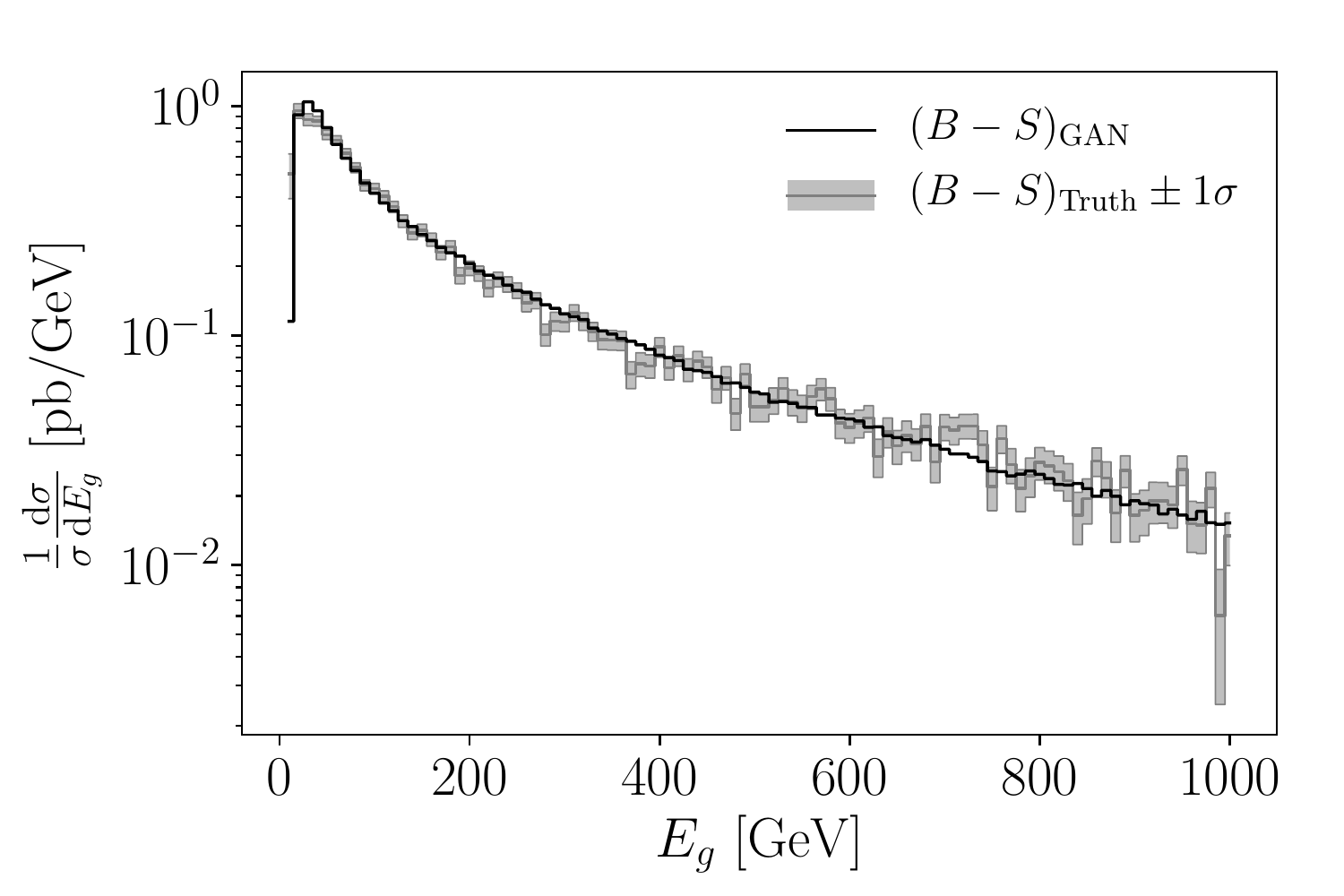}\\
\includegraphics[page=2, width=0.49\textwidth]{figures/cs_dist}
\includegraphics[page=2, width=0.49\textwidth]{figures/cs_dist_zoom}
\caption{Left: Generated (solid) and true (dashed) $Zg$ events at the
  LHC for the two input distributions and the subtracted
  output. Right: distribution of the subtracted events, true and
  generated, including the error envelope propagated from the input
  statistics.}
\label{fig:cs_sub1}
\end{figure}

The problem with this specific process is that the Catani-Seymour
dipoles describe the full matrix element over a huge part of phase
space~\cite{Campbell:2017hsr} and the combination of hard matrix
element and dipoles is typically tiny and negative. We discuss
changing signs in probability distributions in
Sec.~\ref{sec:toy_gen}. In addition, the one distribution a GAN can
never generate is a probability distribution compatible with zero
everywhere. In this case the GAN would either over-fit statistical
fluctuations or become unstable. This is why in our toy application we
shift the Catani-Seymour dipole by a constant such that the
cancellation of the divergent matrix element still works, but the
combined result integrated over phase space remains finite.

Note that the kinematics of our subtraction terms are not the same as
in fixed-order calculations, instead it is similar to the mapping in
the modified subtraction method MC@NLO~\cite{Frixione:2002ik}. 
Within this modified subtraction method the subtraction
events live in the born phase space and thus cannot analytically cancel the real emission contributions
on the integrand level. This means, that both parts need to be evaluated and generated individually and the subtraction
needs to be performed at the level of events. However, regions of large cancellations typically induce large statistical uncertainties owing to binning effects in standard histogram-based methods. In this case the global efficiency of event generators at NLO accuracy is
dominated by the efficiency of computing the subtracted real-emission
corrections, which presents a major challenge for event simulation at
the HL-LHC~\cite{Alves:2017she,Buckley:2019wov}.

The hyper-parameters have to be modified with respect to the
background subtraction example, due to the large cancellations in the
low energy regime. Now, the discriminator and generator networks
consist of 8 layers with 256 and 512 units per layer, respectively. In the generator we alternate LeakyRelu and $\tanh$ activation functions.
We achieve the best and most stable results choosing
$\lambda_{D_i}=10^{-3}$ with a batch size of 1024 events. We train for
60000 epochs, where each epoch consists of 5 iterations in which the
discriminator gets updated twice as much as the generator. In this example, the discriminator gets the events in the $\{E,p_T, \eta, \phi\}$ representation, which is better suited to resolve the $p_T$ distribution. The other
hyper-parameters are kept the same as in the background subtraction.

We show the results from the collinear subtraction in
Fig.~\ref{fig:cs_sub1}. The GAN perfectly reconstructs the
cancellation in the energy spectrum and the transverse momentum of the
emitted gluon. The left panel shows the distribution of the real ($B$)
and dipole ($S$) contributions to the process and their difference
($B-S$). With the logarithmic axis we see that the GAN smoothly
interpolates over the entire energy range. For small gluon energies
and momenta the GAN reproduces the rate increase towards the
(enforced) phase space boundary, including the finite value of the
subtracted combination $B-S$.  Also in the high energy region, which
suffers from low statistics, the GAN nicely matches the truth
distributions. In the right panel we show the subtracted curve as
always including the error envelope of the input data.

As before, we only use the established NLO dipole as a simple
structure to illustrate the features of our subtraction GAN. Proper
applications could be the more complicated subtraction terms beyond
NLO or the subtraction of on-shell
resonances~\cite{GoncalvesNetto:2012yt}. The latter would combine
aspects discussed in Sec.~\ref{sec:lhc_back} and
Sec.~\ref{sec:lhc_coll} and allow for a fully inclusive study of the
kinematics in the off-shell process, without having to actually do a
subtraction and deciding if a given event is more likely to be part
of the on-shell or off-shell sample.

\section{Outlook}

We have shown how to generate events representing the difference
between two input distributions with a GAN. As a toy example we used
events representing a 1-dimensional probability distribution. Because
the GAN interpolates the input while learning the difference between
the two distributions, it circumvents the statistical limitations of
large cancellations. We have found that the GAN-subtracted events lead
to a very stable phase space coverage and beat the statistical
limitations of the input sample over the entire phase space.

For a slightly more realistic setup we have GANned background
subtraction and collinear dipole subtraction for Drell--Yan production
at the LHC. In the first case the network learned on-shell final state
momenta to subtract the photon-induced continuum from the full $e^+
e^-$ production. It could serve as a test case for a background
subtraction for four-body decays, such that the GANned signal events
reflect the kinematic correlations of the actual signal events hidden
in the background.

In the second case we combined the hard matrix element with modified
Catani-Seymour dipoles for gluon emission into a stable finite
prediction of the real emission process.  We are aware of the fact
that our toy examples are not more than an illustration of what a
subtraction GAN can achieve. However, we have shown how to use a GAN
to manipulate event samples avoiding binning (at least in particle
physics) and we hope that some of the people who do LHC event
simulations for a living will find this technique useful.\footnote{As
  always, our data, our codes, and more details are available upon
  request.}\bigskip

\begin{center} \textbf{Acknowledgments} \end{center}

We would like to thank Stefan H\"oche for extremely useful discussions
and for helping us out with specially made \sherpa events on really
short notice.  In addition, we would like to thank Olivier Mattelaer
for his incredibly friendly help with \madgraph. TP would like to
thank Alexander Grohsjean for some very helpful discussions on
possible applications of subtraction GANs.  Finally, we would like to
thank Kirill Melnikov for inspiring this project and Gregor Kasieczka
for his continuous input on machine learning and GANs. RW acknowledges
support by the IMPRS for \textsl{Precision Tests of Fundamental
  Symmetries}.  The research of AB and TP is supported by the Deutsche
Forschungsgemeinschaft (DFG, German Research Foundation) under grant
396021762 --- TRR 257 \textsl{Particle Physics Phenomenology after the
  Higgs Discovery}.

\end{fmffile}

\bibliography{literature}

\providecommand{\href}[2]{#2}\begingroup\raggedright\begin{thebibliography}{10}

\bibitem{Catani:1996jh}
S.~Catani and M.~H. Seymour,
  \href{http://dx.doi.org/10.1016/0370-2693(96)00425-X}{Phys. Lett. {\bfseries
  B378} (1996)  287},
\href{http://arxiv.org/abs/hep-ph/9602277}{{arXiv:hep-ph/9602277 [hep-ph]}}.

\bibitem{Hoche:2018ouj}
S.~Höche, S.~Liebschner, and F.~Siegert,
  \href{http://dx.doi.org/10.1140/epjc/s10052-019-7212-7}{Eur. Phys. J.
  {\bfseries C79} (2019) 9, 728},
\href{http://arxiv.org/abs/1807.04348}{{arXiv:1807.04348 [hep-ph]}}.

\bibitem{GehrmannDeRidder:2005cm}
A.~Gehrmann-De~Ridder, T.~Gehrmann, and E.~W.~N. Glover,
  \href{http://dx.doi.org/10.1088/1126-6708/2005/09/056}{JHEP {\bfseries 09}
  (2005)  056},
\href{http://arxiv.org/abs/hep-ph/0505111}{{arXiv:hep-ph/0505111 [hep-ph]}}.

\bibitem{Frederix:2008hu}
R.~Frederix, T.~Gehrmann, and N.~Greiner,
  \href{http://dx.doi.org/10.1088/1126-6708/2008/09/122}{JHEP {\bfseries 09}
  (2008)  122},
\href{http://arxiv.org/abs/0808.2128}{{arXiv:0808.2128 [hep-ph]}}.

\bibitem{Currie:2013vh}
J.~Currie, E.~W.~N. Glover, and S.~Wells,
  \href{http://dx.doi.org/10.1007/JHEP04(2013)066}{JHEP {\bfseries 04} (2013)
  066},
\href{http://arxiv.org/abs/1301.4693}{{arXiv:1301.4693 [hep-ph]}}.

\bibitem{Catani:2001cc}
S.~Catani, F.~Krauss, R.~Kuhn, and B.~R. Webber,
  \href{http://dx.doi.org/10.1088/1126-6708/2001/11/063}{JHEP {\bfseries 11}
  (2001)  063},
\href{http://arxiv.org/abs/hep-ph/0109231}{{arXiv:hep-ph/0109231 [hep-ph]}}.

\bibitem{Mangano:2002ea}
M.~L. Mangano, M.~Moretti, F.~Piccinini, R.~Pittau, and A.~D. Polosa,
  \href{http://dx.doi.org/10.1088/1126-6708/2003/07/001}{JHEP {\bfseries 07}
  (2003)  001},
\href{http://arxiv.org/abs/hep-ph/0206293}{{arXiv:hep-ph/0206293 [hep-ph]}}.

\bibitem{GoncalvesNetto:2012yt}
D.~Gonçalves-Netto, D.~López-Val, K.~Mawatari, T.~Plehn, and I.~Wigmore,
  \href{http://dx.doi.org/10.1103/PhysRevD.87.014002}{Phys. Rev. {\bfseries
  D87} (2013) 1, 014002},
\href{http://arxiv.org/abs/1211.0286}{{arXiv:1211.0286 [hep-ph]}}.

\bibitem{Plehn:2011nx}
T.~Plehn and M.~Takeuchi,
  \href{http://dx.doi.org/10.1088/0954-3899/38/9/095006}{J. Phys. {\bfseries
  G38} (2011)  095006},
\href{http://arxiv.org/abs/1104.4087}{{arXiv:1104.4087 [hep-ph]}}.

\bibitem{goodfellow}
I.~J. {Goodfellow}, J.~{Pouget-Abadie}, M.~{Mirza}, B.~{Xu}, D.~{Warde-Farley},
  S.~{Ozair}, A.~{Courville}, and Y.~{Bengio},
  \href{http://arxiv.org/abs/1406.2661}{{arXiv:1406.2661 [stat.ML]}}.

\bibitem{bendavid}
J.~Bendavid,
\href{http://arxiv.org/abs/1707.00028}{{arXiv:1707.00028 [hep-ph]}}.

\bibitem{dutch}
S.~Otten {\em et al.},
\href{http://arxiv.org/abs/1901.00875}{{arXiv:1901.00875 [hep-ph]}}.

\bibitem{gan_datasets}
B.~Hashemi, N.~Amin, K.~Datta, D.~Olivito, and M.~Pierini,
  \href{http://arxiv.org/abs/1901.05282}{{arXiv:1901.05282 [hep-ex]}}.

\bibitem{DijetGAN}
R.~Di~Sipio, M.~Faucci~Giannelli, S.~Ketabchi~Haghighat, and S.~Palazzo,
  \href{http://dx.doi.org/10.1007/JHEP08(2019)110}{JHEP {\bfseries 08} (2020)
  110},
\href{http://arxiv.org/abs/1903.02433}{{arXiv:1903.02433 [hep-ex]}}.

\bibitem{Butter:2019cae}
A.~Butter, T.~Plehn, and R.~Winterhalder,
  \href{http://dx.doi.org/10.21468/SciPostPhys.7.6.075}{SciPost Phys.
  {\bfseries 7} (2019)  075},
\href{http://arxiv.org/abs/1907.03764}{{arXiv:1907.03764 [hep-ph]}}.

\bibitem{calogan1}
M.~Paganini, L.~de~Oliveira, and B.~Nachman,
  \href{http://dx.doi.org/10.1103/PhysRevLett.120.042003}{Phys. Rev. Lett.
  {\bfseries 120} (2018) 4, 042003},
\href{http://arxiv.org/abs/1705.02355}{{arXiv:1705.02355 [hep-ex]}}.

\bibitem{calogan2}
M.~Paganini, L.~de~Oliveira, and B.~Nachman,
  \href{http://dx.doi.org/10.1103/PhysRevD.97.014021}{Phys. Rev. {\bfseries
  D97} (2018) 1, 014021},
\href{http://arxiv.org/abs/1712.10321}{{arXiv:1712.10321 [hep-ex]}}.

\bibitem{fast_accurate}
P.~Musella and F.~Pandolfi,
  \href{http://dx.doi.org/10.1007/s41781-018-0015-y}{Comput. Softw. Big Sci.
  {\bfseries 2} (2018) 1, 8},
\href{http://arxiv.org/abs/1805.00850}{{arXiv:1805.00850 [hep-ex]}}.

\bibitem{aachen_wgan1}
M.~Erdmann, L.~Geiger, J.~Glombitza, and D.~Schmidt,
  \href{http://dx.doi.org/10.1007/s41781-018-0008-x}{Comput. Softw. Big Sci.
  {\bfseries 2} (2018) 1, 4},
\href{http://arxiv.org/abs/1802.03325}{{arXiv:1802.03325 [astro-ph.IM]}}.

\bibitem{aachen_wgan2}
M.~Erdmann, J.~Glombitza, and T.~Quast,
  \href{http://dx.doi.org/10.1007/s41781-018-0019-7}{Comput. Softw. Big Sci.
  {\bfseries 3} (2019)  4},
\href{http://arxiv.org/abs/1807.01954}{{arXiv:1807.01954 [physics.ins-det]}}.

\bibitem{ATLASShowerGAN}
ATLAS Collaboration, Tech. Rep. ATL-SOFT-PUB-2018-001, CERN, Geneva,
\newblock \href{http://cds.cern.ch/record/2630433}{Jul, 2018}.

\bibitem{ATLASsimGAN}
ATLAS Collaboration, A.~Ghosh, Tech. Rep. ATL-SOFT-PUB-2019-007, CERN, Geneva,
\newblock \href{http://cds.cern.ch/record/2680531}{Jun, 2019}.

\bibitem{Datta:2018mwd}
K.~Datta, D.~Kar, and D.~Roy,
\href{http://arxiv.org/abs/1806.00433}{{arXiv:1806.00433 [physics.data-an]}}.

\bibitem{shower}
E.~Bothmann and L.~Debbio,
  \href{http://dx.doi.org/10.1007/JHEP01(2019)033}{JHEP {\bfseries 01} (2019)
  033},
\href{http://arxiv.org/abs/1808.07802}{{arXiv:1808.07802 [hep-ph]}}.

\bibitem{locationGAN}
L.~de~Oliveira, M.~Paganini, and B.~Nachman,
  \href{http://dx.doi.org/10.1007/s41781-017-0004-6}{Comput. Softw. Big Sci.
  {\bfseries 1} (2017) 1, 4},
\href{http://arxiv.org/abs/1701.05927}{{arXiv:1701.05927 [stat.ML]}}.

\bibitem{monkshower}
J.~W. Monk, \href{http://dx.doi.org/10.1007/JHEP12(2018)021}{JHEP {\bfseries
  12} (2018)  021},
\href{http://arxiv.org/abs/1807.03685}{{arXiv:1807.03685 [hep-ph]}}.

\bibitem{juniprshower}
A.~Andreassen, I.~Feige, C.~Frye, and M.~D. Schwartz,
  \href{http://dx.doi.org/10.1140/epjc/s10052-019-6607-9}{Eur. Phys. J.
  {\bfseries C79} (2019) 2, 102},
\href{http://arxiv.org/abs/1804.09720}{{arXiv:1804.09720 [hep-ph]}}.

\bibitem{Carrazza:2019cnt}
S.~Carrazza and F.~A. Dreyer,
  \href{http://dx.doi.org/10.1140/epjc/s10052-019-7501-1}{Eur. Phys. J.
  {\bfseries C79} (2019) 11, 979},
\href{http://arxiv.org/abs/1909.01359}{{arXiv:1909.01359 [hep-ph]}}.

\bibitem{Lin:2019htn}
J.~Lin, W.~Bhimji, and B.~Nachman,
  \href{http://dx.doi.org/10.1007/JHEP05(2019)181}{JHEP {\bfseries 05} (2019)
  181},
\href{http://arxiv.org/abs/1903.02556}{{arXiv:1903.02556 [hep-ph]}}.

\bibitem{Bellagente:2019uyp}
M.~Bellagente, A.~Butter, G.~Kasieczka, T.~Plehn, and R.~Winterhalder,
\href{http://arxiv.org/abs/1912.00477}{{arXiv:1912.00477 [hep-ph]}}.

\bibitem{keras}
F.~Chollet. \urlx{https://github.com/fchollet/keras}, 2015.

\bibitem{tensorflow}
M.~Abadi {\em et al.}, CoRR (2016)  ,
  \href{http://arxiv.org/abs/1605.08695}{{arXiv:1605.08695 [cs.DC]}}.

\bibitem{adam}
D.~P. {Kingma} and J.~{Ba},
  \href{http://arxiv.org/abs/1412.6980}{{arXiv:1412.6980 [cs.LG]}}.

\bibitem{madgraph}
J.~Alwall {\em et al.}, \href{http://dx.doi.org/10.1007/JHEP07(2014)079}{JHEP
  {\bfseries 07} (2014)  079},
\href{http://arxiv.org/abs/1405.0301}{{arXiv:1405.0301 [hep-ph]}}.

\bibitem{sPlot}
M.~Pivk and F.~R. Le~Diberder,
  \href{http://dx.doi.org/10.1016/j.nima.2005.08.106}{Nucl. Instrum. Meth. A
  {\bfseries 555} (2005)  356},
  \href{http://arxiv.org/abs/physics/0402083}{{arXiv:physics/0402083}}.

\bibitem{sPlot2}
C.~Langenbruch, \href{http://arxiv.org/abs/1911.01303}{{arXiv:1911.01303
  [physics.data-an]}}.

\bibitem{Bothmann:2019yzt}
E.~Bothmann {\em et al.},
  \href{http://dx.doi.org/10.21468/SciPostPhys.7.3.034}{SciPost Phys.
  {\bfseries 7} (2019) 3, 034},
\href{http://arxiv.org/abs/1905.09127}{{arXiv:1905.09127 [hep-ph]}}.

\bibitem{Campbell:2017hsr}
J.~Campbell, J.~Huston, and F.~Krauss, {\em {The Black Book of Quantum
  Chromodynamics}}.
\newblock Oxford University Press,
\newblock
\href{https://global.oup.com/academic/product/the-black-book-of-quantum-chromodynamics-9780199652747}{2017}.
\newblock

\bibitem{Frixione:2002ik}
S.~Frixione and B.~R. Webber,
  \href{http://dx.doi.org/10.1088/1126-6708/2002/06/029}{JHEP {\bfseries 06}
  (2002)  029},
\href{http://arxiv.org/abs/hep-ph/0204244}{{arXiv:hep-ph/0204244 [hep-ph]}}.

\bibitem{Alves:2017she}
HEP Software Foundation, J.~Albrecht {\em et al.},
  \href{http://dx.doi.org/10.1007/s41781-018-0018-8}{Comput. Softw. Big Sci.
  {\bfseries 3} (2019) 1, 7},
\href{http://arxiv.org/abs/1712.06982}{{arXiv:1712.06982 [physics.comp-ph]}}.

\bibitem{Buckley:2019wov}
A.~Buckley,
\href{http://arxiv.org/abs/1908.00167}{{arXiv:1908.00167 [hep-ph]}}.

\end{thebibliography}\endgroup

\end{document}